\documentclass[twocolumn,aps,pra,showpacs,superscriptaddress]{revtex4}
\usepackage{graphicx}
\usepackage{amsfonts}
\usepackage{amsmath}
\usepackage{amssymb}

\newcommand{\ket}[1]{\left| #1 \right\rangle}
\newcommand{\bra}[1]{\left\langle #1 \right|}
\newcommand{\de}[1]{\left( #1 \right)}

\begin{document}

\title{State-independent contextuality with identical particles}

%%%%%%%%%%%%%%%%%%%%%%%%%%%%%%%%%%%%%%%%%%%%%%%%%%%%%%%%%%%%%%%%%%%

\author{Ad\'an Cabello}
 \email{adan@us.es}
 \affiliation{Departamento de F\'{\i}sica
 Aplicada II, Universidad de Sevilla, E-41012 Sevilla, Spain}
 \affiliation{Universidade Federal de Minas Gerais,
 Caixa Postal 702, 30123-970, Belo Horizonte, MG, Brazil}

\author{Marcelo Terra Cunha}
 \email{tcunha@mat.ufmg.br}
 \affiliation{Departamento de Matem\'atica, Universidade Federal de Minas Gerais,
 Caixa Postal 702, 30123-970, Belo Horizonte, MG, Brazil}

%%%%%%%%%%%%%%%%%%%%%%%%%%%%%%%%%%%%%%%%%%%%%%%%%%%%%%%%%%%%%%%%%%%

\date{\today}

%First version: December 10, 2012 (Belo Horizonte)
%Submitted version: January 03, 2013 (Majadahonda/Belo Horizonte)
%This version: February 25, 2013 (Sevilla). After PRA proofs

%%%%%%%%%%%%%%%%%%%%%%%%%%%%%%%%%%%%%%%%%%%%%%%%%%%%%%%%%%%%%%%%%%%

\begin{abstract}
It has been recently conjectured that the state-independency of quantum contextuality may be lost when the indistinguishability of identical particles is taken into account. Here, we show that quantum state-independent contextuality exists for any system of more than one identical bosonic qudits, and for most systems of fermionic qudits. The only exception is the case of $d$ fermionic qudits, since there the dimension of the antisymmetric subspace is 1, which is insufficient for contextuality. For all the other cases, either the symmetry precludes the existence of physical states, or we provide an explicit method to produce quantum state-independent contextual correlations.
\end{abstract}

%%%%%%%%%%%%%%%%%%%%%%%%%%%%%%%%%%%%%%%%%%%%%%%%%%%%%%%%%%%%%%%%%%%

%Entanglement and quantum nonlocality
%(e.g. EPR paradox, Bell's inequalities, GHZ states, etc.)
%Optical tests of quantum mechanics

\maketitle

%%%%%%%%%%%%%%%%%%%%%%%%%%%%%%%%%%%%%%%%%%%%%%%%%%%%%%%%%%%%%%%%%%%
% Introduction
%%%%%%%%%%%%%%%%%%%%%%%%%%%%%%%%%%%%%%%%%%%%%%%%%%%%%%%%%%%%%%%%%%%

\section{Introduction}

%%%%%%%%%%%%%%%%%%%%%%%%%%%%%%%%%%%%%%%%%%%%%%%%%%%%%%%%%%%%%%%%%%%

All photons are identical in their properties; the same is true of all electrons, all kaons, etc. Particle indistinguishability has important consequences (e.g., it is behind lasers, Bose-Einstein condensation, and superconductivity) and leads to extra postulates in nonrelativistic quantum mechanics (QM). The fact that two physical situations that differ only by the permutation of identical particles do not have any observable difference is called the principle of indistinguishability \cite{MG64}, although it merely defines what we mean by ``identical'' particles. A consequence is that any prediction of QM must be invariant under permutation of identical particles. This leads to a superselection rule prohibiting interference between states of different permutation symmetry. In addition to that, if one keeps the usual QM formulation, relating pure states to one-dimensional projectors (also called rays), there are only two one-dimensional representations of the permutation group \cite{Greenberg09}; the only physical states for identical particles whose spin is integer, called bosons (half-odd integer, called fermions) are the symmetric (antisymmetric) states. Symmetric (antisymmetric) states are those invariant (multiplied by $-1$) under any two-particle interchange \cite{SW00}.

While there are many works investigating entanglement and nonlocality for identical particles \cite{SCKLL01,ESBL02,WV03,SVC04,GM04,CSCLV05,CMMCS07,AFOV08,TMB11}, little is known about how particle indistinguishability and the symmetrization postulate affect quantum contextuality \cite{BFGO11} and state-independent quantum contextuality.

The hypothesis that compatible measurements reveal results that are independent of the choice of which other compatible measurements are jointly performed is inconsistent with QM \cite{Specker60,Bell66,KS67}. This is what is meant by quantum contextuality. Consequently, contextual correlations can be obtained by measuring suitably chosen sets of observables on a system prepared in a suitably chosen quantum state.
A well-known example of quantum contextuality is given by quantum nonlocality, when quantum correlations violate some Bell inequality \cite{Bell64}. Contextual correlations, however, can also be observed on any physical system (not necessarily composite) described in QM by a Hilbert space of dimension $3$ or higher. The importance of having a minimum of three dimensions is explained by the fact that in lower dimensions any observable does not belong to more than one context (i.e., a set of compatible observables), so noncontextuality does not impose any restriction, and the notion of contextuality is meaningless. For example, for two-dimensional Hilbert space, a nontrivial observable $O$ induces a complete splitting of this space as an orthogonal direct sum of two irreducible subspaces, implying that all other observables compatible with this given one are also mutually compatible and all contexts for the observable $O$ are equivalent.

A notable property of quantum contextuality is that the same set of observables can produce contextual correlations for any initial quantum state of the system \cite{Cabello08}. These correlations are called state-independent contextual (SIC) correlations and have been observed in experiments with ions \cite{KZGKGCBR09} and photons \cite{ARBC09,DHANBSC12}. There is a method \cite{BBCP09} for producing SIC correlations from any Kochen-Specker (KS) set of yes-no tests \cite{KS67}. Since explicit KS sets exist for any quantum system with state space of dimension $3$ or higher \cite{CG96,CEG05}, this implies that SIC correlations can be produced for any of these physical systems. The situation does not change by the observation that there are sets of yes-no tests, which are not KS sets but can be used to produce SIC correlations \cite{YO12,BBC12}, since the necessary and sufficient condition for SIC correlations \cite{Cabello12} still requires the quantum systems to have at least dimension~$3$.

In this article, we address the problem of whether SIC correlations can be obtained for physical systems of indistinguishable particles. Recently, Srikanth and Gangopadhyay \cite{SG12} have speculated that the state-independency of quantum contextuality might be lost in this case. Their argument is based of the impossibility of symmetrizing a specific proof of state-independent contextuality for a two-qubit system. The aim of this paper is to show that SIC correlations can be produced with any number ($n \geq 2$) of identical bosonic and most numbers of fermionic qudits, and to provide an explicit method for producing SIC correlations in all these cases.

%%%%%%%%%%%%%%%%%%%%%%%%%%%%%%%%%%%%%%%%%%%%%%%%%%%%%%%%%%%%%%%%%%%
% Proof that SIC correlations exist for most scenarios
%%%%%%%%%%%%%%%%%%%%%%%%%%%%%%%%%%%%%%%%%%%%%%%%%%%%%%%%%%%%%%%%%%%

\section{Proof that SIC correlations exist for most scenarios}

%%%%%%%%%%%%%%%%%%%%%%%%%%%%%%%%%%%%%%%%%%%%%%%%%%%%%%%%%%%%%%%%%%%

Our physical systems consist of $n$ identical fermions (bosons), each of them having $d$ levels, when considered isolated. This $d$ counts all physical degrees of freedom (including the spin levels), otherwise we cannot deal with the symmetrization postulate. A two-ququart fermionic system could be, for example, a system of two spin-$\frac{1}{2}$ particles, each of them with access to two spatial modes. Notice, however, that, with this definition, $n$-qudit {\em fermionic} systems do not exist for odd $d$, since fermions cannot be spinless, and its spin-state space has even dimension. We shall continue the general argument, however, under the assumption that one can, at least in principle, take a fermionic system with two degrees of freedom (e.g., spin and $d$ available spatial modes) and impose a symmetric state of spin, implying an antisymmetric $d$-dimensional effective state space \cite{CSCLV05}. However, a two-qubit fermionic system could be two spin-$\frac{1}{2}$ particles with no extra degrees of freedom (e.g., a doubly occupied quantum dot). For bosons no restriction applies, since we can simply assume that the spin is zero.

Given a Hilbert space ${\cal H} = \bigotimes_{i=1}^n {\cal H}_d$, representing the complete state space of a system of $n$ {\em distinguishable} qudits, we denote by ${\cal S}$ and ${\cal A}$ the totally symmetric and antisymmetric subspaces of ${\cal H}$, respectively. ${\cal S}$ and ${\cal A}$ are mutually orthogonal subspaces and each of them is itself a Hilbert space. The key point for having SIC correlations in identical particles is that each {\em physical} space has to have, at least, dimension $3$. While the dimension of ${\cal H}$ is $d^n$, for the totally symmetric and antisymmetric subspaces we have,
\begin{subequations}
\begin{align}
 {\rm dim}({\cal S})&=\left(\!\!{d \choose n}\!\!\right):={d+n-1 \choose n}, \label{eq:dimS}\\
 {\rm dim}({\cal A})&={d \choose n},\label{eq:dimA}
 \end{align}
 \end{subequations}
the well-known numbers of $n$ combinations of $d$ symbols, allowing for repetitions or not, respectively \cite{JAC04}.
If, for a given system, one of these dimensions is 0, this means that there are no physical states. If it is 3 or greater, then the method described below allows us to obtain SIC correlations. The only physical scenario where SIC correlations are not possible is when one of these dimensions is 1 or 2. A simple inspection shows $2$ never happens, while $1$ only occurs for systems of $d$ fermionic qudits (remember the restrictions already discussed when considering $d$ identical fermionic qudits), i.e., the only physical exception is the generalization to $d$ qudits of the two-qubit singlet.

%%%%%%%%%%%%%%%%%%%%%%%%%%%%%%%%%%%%%%%%%%%%%%%%%%%%%%%%%%%%%%%%%%%
% Method
%%%%%%%%%%%%%%%%%%%%%%%%%%%%%%%%%%%%%%%%%%%%%%%%%%%%%%%%%%%%%%%%%%%

\section{Method}

%%%%%%%%%%%%%%%%%%%%%%%%%%%%%%%%%%%%%%%%%%%%%%%%%%%%%%%%%%%%%%%%%%%

A method for revealing SIC correlations using a bosonic (fermionic) system has three steps: (i) Choose an orthogonal basis $B$ of ${\cal S}$ (${\cal A}$). (ii) Choose a KS set in dimension ${\rm dim}({\cal S})$ [${\rm dim}({\cal A})$] and rewrite it in terms of $B$: this produces a KS set in ${\cal S}$ (${\cal A}$) containing only symmetric (antisymmetric) rank-one projectors. (iii) The final step consists in applying the method in Ref.~\cite{BBCP09} to, given a KS set, construct a noncontextuality inequality (i.e., satisfied by any theory that assumes that results of tests are independent on whether or not other compatible tests are performed) violated by any state in this space. In this case, a state-independent noncontextuality inequality violated by any symmetric (antisymmetric) quantum state, i.e., revealing SIC correlations in bosonic (fermionic) systems.

We illustrate steps (i) and (ii) with two examples. After that, step (iii) is a direct application of the result in Ref.~\cite{BBCP09}, which we make explicit here only in the second example.

Consider two {\em bosonic} qutrits. In this case, Eq.~(\ref{eq:dimS}) indicates that ${\rm dim}({\cal S})=6$. An orthonormal basis of completely symmetric states is the following one (in obvious notation):
\begin{subequations}
\begin{align}
 \ket{\hat{0}} & = \ket{++},\\
 \ket{\hat{1}} & = \frac{1}{\sqrt{2}}\left(\ket{+0}+\ket{0+}\right),\\
 \ket{\hat{2}} & = \frac{1}{\sqrt{6}}\left(\ket{+-}+2 \ket{00} +\ket{-+}\right),\\
 \ket{\hat{3}} & = \frac{1}{\sqrt{2}}\left(\ket{0-}+\ket{-0}\right),\\
 \ket{\hat{4}} & = \ket{--},\\
 \ket{\hat{5}} & = \frac{1}{\sqrt{3}}\left(\ket{+-} - \ket{00} +\ket{-+}\right).
\end{align}
\end{subequations}
Using this basis and results in Ref.~\cite{CEG05}, we can construct a KS set in dimension 6, containing only rank-one projectors onto the following 31 symmetric vectors:
$S_6:=\{({\bf a},0,0)$, $(0,0,{\bf a}):{\bf a} \in S_4\}$ $\cup$
$\{(0,1,0,0,0,0)$, $(1,0,-1,0,0,0)$, $(1,1,1,1,0,0)$\}
$-$ $\{(0,0,1,0,0,0)$, $(0,0,0,1,0,0)$, $(1,1,0,0,0,0)$,
$(0,0,1,-1,0,0)$, \linebreak $(1,-1,-1,1,0,0)$, $(0,1,0,1,0,0)\}$, where $S_4$ is the following KS set in dimension 4 \cite{CEG96}:
$S_4:=\{(1,0,0,0)$, $(0,0,1,0)$,
$(0,0,0,1)$, $(1,1,0,0)$, $(0,1,1,0)$, $(0,0,1,1)$, $(1,-1,0,0)$,
$(0,1,-1,0)$, $(1,0,1,0)$, $(0,1,0,1)$, $(0,1,0,-1)$, $(1,0,0,1)$,
$(1,-1,1,-1)$, $(1,1,-1,-1)$, \linebreak $(1,-1,-1,1)$,
$(1,1,1,-1)$, $(1,1,-1,1)$, $(-1,1,1,1)\}$.

In the case of two {\em fermionic} qutrits, Eq.~(\ref{eq:dimA}) indicates that ${\rm dim}({\cal A})=3$. A basis of completely antisymmetric states is the following one:
\begin{subequations}
 \label{eq:fermBasis}
\begin{align}
 \ket{\tilde{0}} & = \frac{1}{\sqrt{2}}\left(\ket{+0}-\ket{0+}\right),\\
 \ket{\tilde{1}} & = \frac{1}{\sqrt{2}}\left(\ket{+-}-\ket{-+}\right),\\
 \ket{\tilde{2}} & = \frac{1}{\sqrt{2}}\left(\ket{0-}-\ket{-0}\right).
\end{align}
\end{subequations}
Using this basis and results in Ref.~\cite{Peres95}, we can construct a KS set in dimension 3, containing only rank-one projectors onto the following 31 antisymmetric vectors $A_3:=\{P(0,0,1), P(0,1,1), P(0,1,-1), P(0,1,2), P(0,1,-2)$, \linebreak $(1,1,1), P(1,1,-1), P(1,1,2), P(1,1,-2), P(1,-1,2)\}$ $-$ $\{(2,1,1), (2,1,0), (2,1,-1), (-1,2,1), (1,-2,0), (1,-2,1)\}$, where $P(a,b,c)$ is the set of all vectors with components $a$, $b$, $c$, and representing {\em different} directions. For step (iii), it is important to recognize that these $31$ vectors include $17$ triorthogonal frames (see the Appendix), which implies \cite{BBCP09} the noncontextuality inequality
\begin{subequations}
\begin{equation}
\beta (3,17) \leq 15,
\end{equation}
where
\begin{equation}
\beta (3,17) = \sum _{j=1}^{17} \left\langle B^j \right\rangle ,
\end{equation}
\begin{equation}\label{eq:Bj}
B^j = - \left(
1 + A_1^jA_2^j + A_2^jA_3^j + A_3^jA_1^j + A_1^jA_2^jA_3^j
\right),
\end{equation}
\end{subequations}
and each $A_i^j$ is considered as a noncontextual random variable assuming the values $\pm 1$, in the sense that when the same random variable belongs to more than one context, it must be assigned the same value.
The contradiction with quantum mechanics is obtained when we make $A_i^j = \mathbb{I} - 2\ket{v_i^j}\bra{v_i^j}$ and we promptly verify that
\begin{equation}
\beta _{\text{QM}} = 17,
\end{equation}
irrespectively of the quantum state considered.

%%%%%%%%%%%%%%%%%%%%%%%%%%%%%%%%%%%%%%%%%%%%%%%%%%%%%%%%%%%%%%%%%%%
% Conclusions
%%%%%%%%%%%%%%%%%%%%%%%%%%%%%%%%%%%%%%%%%%%%%%%%%%%%%%%%%%%%%%%%%%%

\section{Conclusions}

%%%%%%%%%%%%%%%%%%%%%%%%%%%%%%%%%%%%%%%%%%%%%%%%%%%%%%%%%%%%%%%%%%%

We show here that it is not true that particle indistinguishability flaws state-independent contextuality. Indeed, for $n$ qudits we have the following situation: for $n\geq 2$ bosonic qudits, $d \geq 2$, the symmetric subspace $\mathcal{S}$ has dimension $\left(\!{n \choose d}\!\right) \geq 3$ and one can always find a KS set of vectors suitable for generating a noncontextuality inequality violated by any physical state.

The fermionic case is richer. First of all, if we consider the qudit as the complete description of an isolated fermion, we need $d$ to be even. But even if we considered effective spaces with fermionic symmetry, since the antisymmetric subspace $\mathcal{A}$ has dimension ${n \choose d}$, one would need $n \geq d$ for the fermionic subspace to exist. For SIC to exist with $n$ fermionic qudits ($d \geq 2$), it is sufficient to have $n > d$.

The distinctive characteristic of nonlocality among other forms of contextuality is the requirement that particles are spacelike separated. If we remove this requirement, we end up with general contextuality. Here, we have taken a further step and addressed the problem of what happens when we also remove the assumption of particle distinguishability. To a certain extent, this means investigating quantum contextuality in the extreme opposite to where it is better understood and tested (namely, nonlocal scenarios). The result in this work shows that quantum contextuality, in its more powerful version, remains, and that, consequently, there is an interesting road for experimental research at a fundamental level: the design and performance of experiments where quantum contextuality can be checked for identical indistinguishable particles, and in which both the state preparation and measurements are done in a regime where it is impossible to label the particles.

%%%%%%%%%%%%%%%%%%%%%%%%%%%%%%%%%%%%%%%%%%%%%%%%%%%%%%%%%%%%%%%%%%%
% Acknowledgments
%%%%%%%%%%%%%%%%%%%%%%%%%%%%%%%%%%%%%%%%%%%%%%%%%%%%%%%%%%%%%%%%%%%

\begin{acknowledgments}
This work was supported by the Spanish Ministry of Economy and Competitiveness through Project No.\ FIS2011-29400, the Brazilian program Science without Borders, the Brazilian National Institute for Science and Technology of Quantum Information, and the Brazilian agencies Capes, CNPq, and Fapemig.
\end{acknowledgments}

\appendix

%%%%%%%%%%%%%%%%%%%%%%%%%%%%%%%%%%%%%%%%%%%%%%%%%%%%%%%%%%%%%%%%%%%
% Appendix
%%%%%%%%%%%%%%%%%%%%%%%%%%%%%%%%%%%%%%%%%%%%%%%%%%%%%%%%%%%%%%%%%%%

\section{}

%%%%%%%%%%%%%%%%%%%%%%%%%%%%%%%%%%%%%%%%%%%%%%%%%%%%%%%%%%%%%%%%%%%

We label the $17$ triorthogonal frames obtained from the $31$ vectors of KS proof in Ref.~\cite{Peres95} as
\begin{subequations}
\begin{align}
\de{v_1^1,v_2^1,v_3^1}&= \de{\de{1,0,0},\de{0,1,0},\de{0,0,1}}, \\
\de{v_1^2,v_2^2,v_3^2}&= \de{\de{1,0,0},\de{0,1,1},\de{0,1,-1}}, \\
\de{v_1^3,v_2^3,v_3^3}&= \de{\de{1,0,1},\de{0,1,0},\de{-1,0,1}}, \\
\de{v_1^4,v_2^4,v_3^4}&= \de{\de{1,1,0},\de{1,-1,0},\de{0,0,1}}, \\
\de{v_1^5,v_2^5,v_3^5}&= \de{\de{1,0,0},\de{0,1,2},\de{0,-2,1}}, \\
\de{v_1^6,v_2^6,v_3^6}&= \de{\de{1,0,0},\de{0,1,-2},\de{0,2,1}}, \\
\de{v_1^7,v_2^7,v_3^7}&= \de{\de{1,0,2},\de{0,1,0},\de{-2,0,1}}, \\
\de{v_1^8,v_2^8,v_3^8}&= \de{\de{1,0,-2},\de{0,1,0},\de{2,0,1}}, \\
\de{v_1^9,v_2^9,v_3^9}&= \de{\de{1,2,0},\de{-2,1,0},\de{0,0,1}}, \\
\de{v_1^{10},v_2^{10},v_3^{10}}&= \de{\de{1,1,1},\de{1,-1,0},\de{1,1,-2}}, \\
\de{v_1^{11},v_2^{11},v_3^{11}}&= \de{\de{1,1,1},\de{0,1,-1},\de{-2,1,1}}, \\
\de{v_1^{12},v_2^{12},v_3^{12}}&= \de{\de{1,1,-1},\de{0,1,1},\de{2,-1,1}}, \\
\de{v_1^{13},v_2^{13},v_3^{13}}&= \de{\de{1,-1,1},\de{1,1,0},\de{-1,1,2}}, \\
\de{v_1^{14},v_2^{14},v_3^{14}}&= \de{\de{-1,1,1},\de{1,0,1},\de{1,2,-1}}, \\
\de{v_1^{15},v_2^{15},v_3^{15}}&= \de{\de{-1,1,1},\de{1,1,0},\de{1,-1,2}}, \\
\de{v_1^{16},v_2^{16},v_3^{16}}&= \de{\de{1,1,-1},\de{1,-1,0},\de{1,1,2}}, \\
\de{v_1^{17},v_2^{17},v_3^{17}}&= \de{\de{1,-1,1},\de{-1,0,1},\de{1,2,1}}.
\end{align}
\end{subequations}
It is essential to note the multiple labeling of the vectors; e.g., $v_1^1 = v_1^2 = v_1^5 = v_1^6$, as well as $v_3^2 = v_2^{11}$, and so on, which demands that $A_1^1 = A_1^2 = A_1^5 = A_1^6$, as well as $A_3^2 = A_2^{11}$, and so on (this is the noncontextuality hypothesis). It is also essential that this set of vectors gives a KS proof, in the sense that it is impossible to make a noncontextual assignation of values $\pm 1$ to the variables $A_i^j$ also obeying that, in each context, one and only one of the variables receives the value $-1$ [and this is clearly the condition for each $B^j$ be maximal; see Eq.~\eqref{eq:Bj}]. This makes the noncontextual bound to be two units less than the algebraic bound, obtained by the independent maximization of each $B^j$. Finally, in our case one must remember that these vectors are all written with respect to the basis $\left\{ \ket{\tilde{0}}, \ket{\tilde{1}}, \ket{\tilde{2}} \right\}$ of Eqs.~\eqref{eq:fermBasis}, which makes the quantum operators to be
\begin{subequations}
\begin{align}
A_1^1 = A_1^2 = A_1^5 = A_1^6 =& \mathbb{I} - 2 \ket{\tilde{0}}\bra{\tilde{0}},\\
A_2^1 = A_2^3 = A_2^7 = A_2^8 =& \mathbb{I} - 2 \ket{\tilde{1}}\bra{\tilde{1}},\\
A_3^1 = A_3^4 = A_3^9 =& \mathbb{I} - 2 \ket{\tilde{2}}\bra{\tilde{2}},
\end{align}
\end{subequations}
and so on.

%%%%%%%%%%%%%%%%%%%%%%%%%%%%%%%%%%%%%%%%%%%%%%%%%%%%%%%%%%%%%%%%%%%

%%%%%%%%%%%%%%%%%%%%%%%%%%%%%%%%%%%%%%%%%%%%%%%%%%%%%%%%%%%%%%%%%%%


\begin{thebibliography}{99}

%%%%%%%%%%%%%%%%%%%%%%%%%%%%%%%%%%%%%%%%%%%%%%%%%%%%%%%%%%%%%%%%%%%

\bibitem{MG64}
 A.M.L. Messiah and O.W. Greenberg,
 %Symmetrization postulate and its experimental foundation
 Phys. Rev. \textbf{136}, B 248 (1964).

 \bibitem{Greenberg09}
 O.W. Greenberg,
 %``Generalizations of quantum statistics,''
 in {\em Compendium of Quantum Physics},
 edited by D. Greenberger, K. Hentschel, and F. Weinert
 (Springer, Berlin, 2009), p.~255.
 %255-258

\bibitem{SW00}
 R.F. Streater and A.S. Wightman,
 {\em PCT, Spin and Statistics, and All That}
 (Princeton University Press, Princeton, NJ, 2000).

%%%%%%%%%%%%%%%%%%%%%%%%%%%%%%%%%%%%%%%%%%%%%%%%%%%%%%%%%%%%%%%%%%%

\bibitem{SCKLL01}
 J. Schliemann, J.I. Cirac, M. Ku\'s, M. Lewenstein, and D. Loss,
 %Quantum correlations in two-fermion systems.
 Phys. Rev. A \textbf{64}, 022303 (2001).

\bibitem{ESBL02}
 K. Eckert, J. Schliemann, D. Bru\ss, and M. Lewenstein,
 %Quantum Correlations in Systems of Indistinguishable Particles.
 Ann. Phys. (N.Y.) \textbf{299}, 88 (2002).

\bibitem{WV03}
 H.M. Wiseman and J.A. Vaccaro,
 %Entanglement of indistinguishable particles shared between two parties
 Phys. Rev. Lett. \textbf{91}, 097902 (2003).

\bibitem{SVC04}
 N. Schuch, F. Verstraete, and J.I. Cirac,
 %Nonlocal resources in the presence of superselection rules.
 Phys. Rev. Lett. \textbf{92}, 087904 (2004).

\bibitem{GM04}
 G.-C. Ghirardi and L. Marinatto,
 %General criterion for the entanglement of two indistinguishable particles.
 Phys. Rev. A \textbf{70}, 012109 (2004).

\bibitem{CSCLV05}
 D. Cavalcanti, M. Fran\c ca Santos, M.O. Terra Cunha, C. Lunkes, and V. Vedral,
 %Increasing identical particle entanglement by fuzzy measurements
 Phys. Rev. A \textbf{72}, 062307 (2005).

\bibitem{CMMCS07}
 D. Cavalcanti, L.M. Malard, F.M. Matinaga, M.O. Terra Cunha, and M. Fran\c ca Santos,
 %Useful entanglement from the Pauli principle.
 Phys. Rev. B \textbf{76}, 113304 (2007).

\bibitem{AFOV08}
 L. Amico, R. Fazio, A. Osterloh, and V. Vedral,
 %Entanglement in many-body systems.
 Rev. Mod. Phys. \textbf{80}, 517 (2008).

\bibitem{TMB11}
 M.C. Tichy, F. Mintert, and A. Buchleitner,
 %Essential entanglement for atomic and molecular physics.
 J. Phys. B \textbf{44}, 192001 (2011).

\bibitem{BFGO11}
F. Benatti, R. Floreanini, M. Genovese, and S. Olivares,
%Quantum contextuality in $N$-boson systems
Phys. Rev. A \textbf{84}, 034102 (2011).

%%%%%%%%%%%%%%%%%%%%%%%%%%%%%%%%%%%%%%%%%%%%%%%%%%%%%%%%%%%%%%%%%%%

\bibitem{Specker60}
 E.P. Specker,
 %Die Logik nicht gleichzeitig entscheidbarer Aussagen.
 Dialectica \textbf{14}, 239 (1960).

\bibitem{Bell66}
 J.S. Bell,
 %On the problem of hidden variables in quantum mechanics.
 Rev. Mod. Phys. \textbf{38}, 447 (1966).

\bibitem{KS67}
 S. Kochen and E.P. Specker,
 %The problem of hidden variables in quantum mechanics.
 J. Math. Mech. \textbf{17}, 59 (1967).

\bibitem{Bell64}
 J.S. Bell,
 %``On the Einstein-Podolsky-Rosen paradox'',
 Physics \textbf{1}, 195 (1964).

\bibitem{Cabello08}
 A. Cabello,
 %``Experimentally testable state-independent quantum contextuality'',
 Phys. Rev. Lett. \textbf{101}, 210401 (2008).

\bibitem{KZGKGCBR09}
 G. Kirchmair, F. Z\"ahringer, R. Gerritsma, M. Kleinmann,
 O. G{\"u}hne, A. Cabello, R. Blatt, and C.F. Roos,
 %State-independent experimental test of quantum contextuality.
 Nature (London) \textbf{460}, 494 (2009).

\bibitem{ARBC09}
 E. Amselem, M. R{\aa }dmark, M. Bourennane, and A. Cabello,
 %State-independent quantum contextuality with single photons.
 Phys. Rev. Lett. \textbf{103}, 160405 (2009).

\bibitem{DHANBSC12}
 V. D'Ambrosio, I. Herbauts, E. Amselem, E. Nagali, M. Bourennane, F. Sciarrino, and A. Cabello,
 %Experimental implementation of a Kochen-Specker set of quantum tests
 Phys. Rev. X \textbf{3}, 011012 (2013).

\bibitem{BBCP09}
 P. Badzi{\c a}g, I. Bengtsson, A. Cabello, and I. Pitowsky,
 %``Universality of state-independent violation of inequalities for non-contextual
 %theories'',
 Phys. Rev. Lett. \textbf{103}, 050401 (2009).

\bibitem{CG96}
 A. Cabello and G. Garc\'{\i}a-Alcaine,
 %``Bell-Kochen-Specker theorem for any finite dimensions $n \ge 3$'',
 J. Phys. A \textbf{29}, 1025 (1996).

\bibitem{CEG05}
 A. Cabello, J.M. Estebaranz, and G. Garc\'{\i}a-Alcaine,
 %``Recursive proof of the Bell-Kochen-Specker theorem in any dimension $n \ge 3$'',
 Phys. Lett. A \textbf{339}, 425 (2005).

\bibitem{YO12}
 S. Yu and C.H. Oh,
 %``State-independent proof of Kochen-Specker theorem with 13 rays'',
 Phys. Rev. Lett. \textbf{108}, 030402 (2012).

\bibitem{BBC12}
 I. Bengtsson, K. Blanchfield, and A. Cabello,
 %``A Kochen-Specker inequality from a SIC'',
 Phys.~Lett.~A \textbf{376}, 374 (2012).

\bibitem{Cabello12}
 A. Cabello,
 \eprint{arXiv:1112.5149}.

%%%%%%%%%%%%%%%%%%%%%%%%%%%%%%%%%%%%%%%%%%%%%%%%%%%%%%%%%%%%%%%%%%%

\bibitem{SG12}
 R. Srikanth and D. Gangopadhyay,
% ``Bell-Kochen-Specker theorems for indistinguishable particles''
 unpublished.
%See also \eprint{arXiv:1201.5080}.

%%%%%%%%%%%%%%%%%%%%%%%%%%%%%%%%%%%%%%%%%%%%%%%%%%%%%%%%%%%%%%%%%%%

\bibitem{JAC04}
 I. Jexa, E. Andersson, and A. Chefles,
 %Comparing the states of many quantum systems
 J. Mod. Opt. \textbf{51}, 505 (2004).

%%%%%%%%%%%%%%%%%%%%%%%%%%%%%%%%%%%%%%%%%%%%%%%%%%%%%%%%%%%%%%%%%%%

\bibitem{CEG96}
 A. Cabello, J.M.~Estebaranz, and G.~Garc\'\i a-Alcaine,
 %Bell-Kochen-Specker theorem: A proof with 18 vectors
 Phys. Lett.~A \textbf{212}, 183 (1996).

\bibitem{Peres95}
 A. Peres,
 {\em Quantum Theory: Concepts and Methods}
 (Kluwer, Dordrecht, 1995).

%%%%%%%%%%%%%%%%%%%%%%%%%%%%%%%%%%%%%%%%%%%%%%%%%%%%%%%%%%%%%%%%%%%

\end{thebibliography}
\end{document}